# Stochastic Modeling Approaches for Analyzing Blockchain: A Survey

Hongyue Kang[a*], Xiaolin Chang[a*], Jelena Mišić[b+], Vojislav B. Mišić[b+], Yingying Yao[a*], Zhi Chen[a*]
[a]Beijing Key Laboratory of Security and Privacy in Intelligent Transportation, Beijing Jiaotong University, P. R. China
[b]Ryerson University, Toronto, ON, Canada
Email: *{19112051, xlchang, 17112100, chenzhi}@bjtu.edu.cn; +{ jmisic, vmisic}@ryerson.ca

*Abstract*—Blockchain technology has been attracting much attention from both academia and industry. It brings many benefits to various applications like Internet of Things. However, there are critical issues to be addressed before its widespread deployment, such as transaction efficiency, bandwidth bottleneck, and security. Techniques are being explored to tackle these issues. Stochastic modeling, as one of these techniques, has been applied to analyze a variety of blockchain characteristics, but there is a lack of a comprehensive survey on it. In this survey, we aim to fill the gap and review the stochastic models proposed to address common issues in blockchain. Firstly, this paper provides the basic knowledge of blockchain technology and stochastic models. Then, according to different objects, the stochastic models for blockchain analysis are divided into network-oriented and application-oriented (mainly refer to cryptocurrency). The network-oriented stochastic models are further classified into two categories, namely, performance and security. About the application-oriented stochastic models, the widest adoption mainly concentrates on the price prediction of cryptocurrency. Moreover, we provide analysis and comparison in detail on every taxonomy and discuss the strengths and weaknesses of the related works to serve guides for further researches. Finally, challenges and future research directions are given to apply stochastic modeling approaches to study blockchain. By analyzing and classifying the existing researches, we hope that our survey can provide suggestions for the researchers who are interested in blockchain and good at using stochastic models as a tool to address problems.

## 1. INTRODUCTION

Blockchain is a kind of combined technology, which at least includes cryptography, mathematics, algorithm, and economic models. It is in fact a distributed and open ledger running over a peer-to-peer (P2P) network that can manage transactions for multiple entities efficiently without a middleman. Blockchain was originally created to support the famous cryptocurrency Bitcoin, introduced by Nakamoto [1] in 2008. Since then, many other blockchain platforms (Ethereum [2], Hyperledger Fabric [3], Ripple [4], etc.) have been inspired by Bitcoin and their applications have spread to Internet of Things (IoT) [5], insurance [6], medical science [7] and so on.

Although blockchain has a huge prospect in the future, some concerns have to be considered and tackled. For instance, Coinrail and Bithumb [8], South Korean cryptocurrency exchanges, were hacked in June 2018. In a very short period of time, two hacker attacks took place, which resulted in 5300 Bitcoins ($ 40 million) and $31 million losses respectively. Therefore, security problems of blockchain need to be solved. Another concern is that the performance of blockchain needs to be improved. For example, the transactions per second (TPS) of blockchain platforms is several orders of magnitude lower [9] than that of traditional database systems like VISA and Paypal [10].

Over the past few years, researchers have adopted many techniques to tackle these issues. Among these techniques, stochastic modeling has been employed as a solution in blockchain. A stochastic model forecasts the probability of various outcomes under different conditions, using random variables [11]. Thus, stochastic models can be used to obtain the probability distributions of many metrics in blockchain networks. Through the stochastic models analysis, the potential relationship among metrics can be proved. Moreover, stochastic models can be utilized to learn and predict miners' behaviors and then the optimal strategy is chosen to get more rewards. In addition, stochastic models have made many contributions to blockchain in many aspects, like evaluating reliability [12], analyzing performance [13] and achieving traceability [14]. Therefore, stochastic models are natural considerations for analyzing blockchain.

There are several good survey papers on blockchain. Some survey papers focus on the specific aspects of blockchain such as consensus protocols [15], privacy, security [16], smart contracts [17], and so on. Some papers concentrate on surveying the applications of blockchain, such as IoT [18] and industry [19]. However, there is no survey to conclude the situation where they use stochastic models to explore blockchain. Moreover, it is important for researchers to know the open challenges and research trend about stochastic models used to address blockchain issues. Motivated by it, we aim to present the survey with the comprehensive literate review on the stochastic models in blockchain. We hope that our effort could give references for people who want to study blockchain using stochastic models. Most of the related works are after 2016 and based on the different objects, the stochastic models for analyzing blockchain are divided into network-oriented and application-oriented (mainly refer to cryptocurrency). The network-oriented stochastic models are classified into two categories, namely, performance and security. About the blockchain application-oriented stochastic models, the widest adoption mainly concentrates on the price prediction of cryptocurrency. The detailed classification of our paper is shown in Fig. 1. The contributions of our survey are as follows:

(1) We give an overview of blockchain, present several common stochastic models, and discuss what kinds of problems in blockchain are the stochastic models good at solving.

(2) We compare the existing works which apply stochastic models to study blockchain attributes, and analyze their

strengths and weaknesses in order to serve guides for further researches.

(3) We conclude open research challenges in combining blockchain and stochastic models. In addition, we point out several future directions of using stochastic models to solve blockchain issues.

The paper is organized as follows. Section 2 gives the overview and fundamentals of blockchain and stochastic models. Section 3 introduces the stochastic models for blockchain networks performance. Section 4 discusses the stochastic models for security in blockchain networks. Section 5 presents stochastic models for blockchain applications (mainly refer to cryptocurrency). Section 6 outlines challenges and future research directions. The conclusion is given in Section 7.

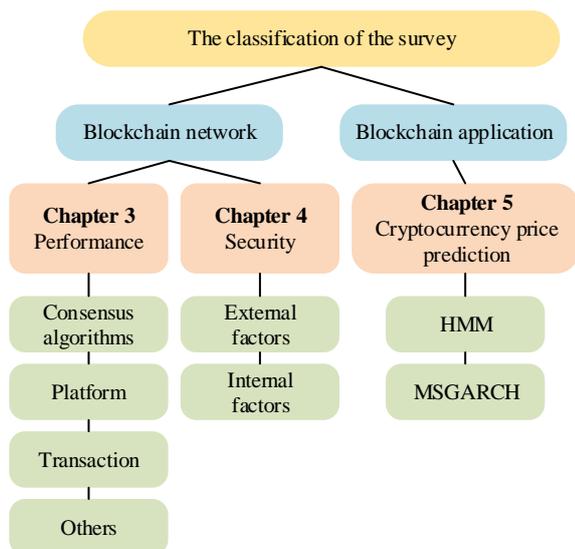

Fig. 1.  The classification of the paper.

## 2. OVERVIEW AND PRELIMINARY

In this section, we give an overview of blockchain on its concepts and give a brief introduction of stochastic models.

### 2.1 Overview of Blockchain

The traditional payment method needs a trusted third party, but the middleman is untrustworthy, leading to various security problems. The emergence of blockchain can create trust among participants without the need for third parties. It has got a lot of attention for the following reasons:

- Greater transparency: In addition to the identity information of the transaction parties being encrypted, the data is transparent to the nodes of the whole blockchain network. Anyone or participating nodes can query blockchain data records or develop relevant applications through open interfaces, which is the basis of blockchain systems to be trusted. Therefore, data on blockchain has high transparency.

- Enhanced security: If anyone wants to change the information in blockchain, more than half of the total hash power must be controlled to change the data, which is very difficult. In distributed networks, security issues can be better addressed without the trust doubt of third parties.

- Improved traceability: The next block has a hash value of the previous block as a hook. Only the previous hash value is recognized, can the block be hung. Because the block is uniquely identified, it is convenient to query the date that we need.

- Increased efficiency and speed: Because each node in blockchain networks has the right to record without the check by a middleman, which leads to high efficiency and speed.

- Reduced costs: In blockchain, no third party or middleman is required, which saves a lot of labor costs.

### 2.2 Type of blockchain

According to the access mechanisms, the blockchain technologies can be divided into three types: Public blockchain, Private blockchain, and Consortium blockchain [2].

In public blockchain, any individual or group can send transactions or develop their own applications on the public blockchain. Public blockchain is completely decentralized. The access threshold is low in public blockchain networks. Anyone who has a computer and can connect to the Internet can access it. All data is open to the participants. Bitcoin [20] and Ethereum [21] are both public blockchain platforms. Fig. 2 (a) shows an example of public blockchain.

In private blockchain, write permission is controlled by an organization or institution. The privilege of participant nodes will be strictly restricted. Private blockchain is more flexible than public blockchain, which is widely used in enterprises. Hyperledger Fabric and R3 Corda are private blockchain projects [22][23]. Fig. 2 (b) represents a private blockchain.

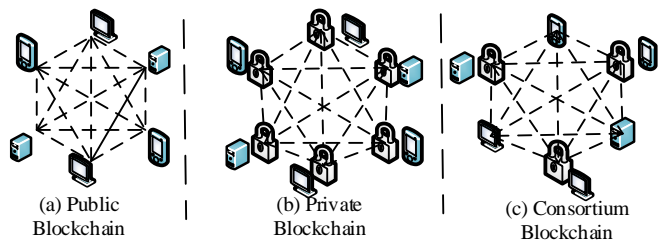

(a) Public Blockchain | (b) Private Blockchain | (c) Consortium Blockchain

Fig. 2.  Types of Blockchain.

In consortium blockchain, it is partially private. Different from private blockchain, consortium blockchain is managed by a group rather than a single entity. Consortium blockchain needs to choose several nodes as record nodes in advance. The generation of each block is determined by the record nodes. Other nodes can trade, but have no right of record. Hyperledger [24] and R3CEV [25] are both consortium blockchain platforms. Fig. 2 (c) gives an example of consortium blockchain. Both consortium blockchain and private blockchain belong to permissioned blockchain. Hyperledger Fabric is a popular permissioned blockchain platform.

### 2.3 Data organization

The data organization of the blockchain can be divided into three levels: transaction, block and chain. Each level has different components to protect data integrity and authenticity.

- Transaction: Transaction is the most basic data structure of blockchain. Block is a data structure for recording transactions, reflecting the fund flow of a transaction. Actually, a transaction is a collection of inputs and outputs, which identifies the senders and receivers and includes the token value/states and some transaction fees [26].

- Block: A block consists of a block header (block hash) to identify itself, a hash pointer [27] to point the previous block, and transactions. The transactions exist in a block as a Merkle root for lightweight storage. The Merkle root is the root of Merkle tree [28]. A Merkel tree is a transaction set organized by a binary tree. Each leaf node represents a transaction labeled with hashcode and the non-leaf node is labeled with the hashcode of the concatenated labels of its two children. This kind of structure is convenient to check the content and increases the difficulty of tampering the data in a block without being noticed.

- Chain: Every block has the previous hash value, so blocks could be chained together, thus forming a "chain". The first block is called genesis block [29]. An example of blockchain is shown in Fig. 3.

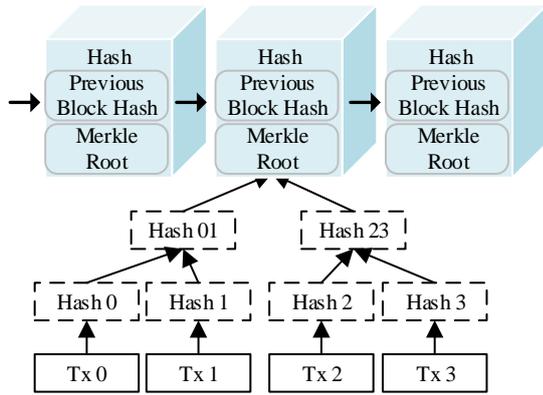

Fig. 3. An example of blockchain.

### 2.4 Consensus

We know that there is no central authority in blockchain, and consensus is the core part of blockchain networks to guarantee transactions to be secured and verified [30]. Generally, it is a technique to reach an agreement in a group. Consensus algorithms can be divided into two types: proof-based and voting-based [31].

In proof-based consensus algorithms, any node must solve a difficult mathematical puzzle if it pretends to produce a new block and add it to the blockchain. The process of getting the right to append the block is called mining and the nodes are called miners. In public blockchain, nodes get rewards after mining successfully, which encourages more miners to participate in blockchain networks. There are different versions of proof-based consensus algorithms, such as Proof of Elapsed time, Proof of Work (PoW), Proof of Stake (PoS) [32], and their hybrid versions. However, PoW needs to consume high electricity. A validator is chosen to generate a new block based on its economic state. The more mining machines you have, the more likely you are to succeed. And Pow is vulnerable to 51%-attacks [33]. In order to overcome high electricity consumption, Proof-of-Stake (PoS) is developed [34]. It is similar to the shareholder mechanism in real life. The more shares people own, the easier it is to get the right to record.

Voting-based consensus algorithms are popular in private blockchain where the nodes are identified. Voting-based consensus algorithms allow nodes to join and leave from the checking system freely [35]. Byzantine mechanism is a typical voting-based consensus algorithm. The most famous Byzantine algorithm is Practical Byzantine Fault Tolerance (PBFT) introduced in the late 90s [36]. It is designed to work in asynchronous systems and aims to reduce the influence of the faulty nodes. Specifically, there are $n$ nodes and $f$ is the number of Byzantine faulty nodes. When $n \geq 3 \times f + 1$, the consensus can be reached in PBFT.

### 2.5 How blockchain works

The main working process of blockchain is as follows (see Fig. 4):

1) User requests a transaction;

2) The transaction is included in the block;

3) The block is broadcast to other nodes in the blockchain network;

4) Nodes validate the transaction;

5) A new block of data is created and appended to blockchain;

6) The transaction is completed.

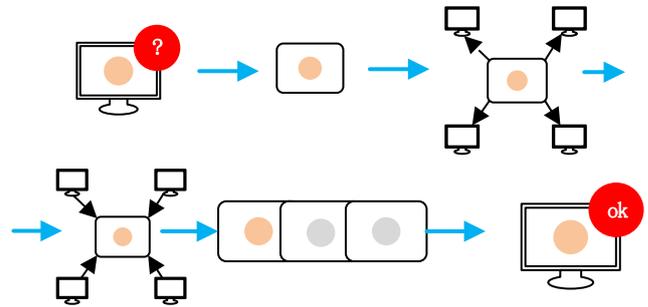

Fig. 4. Blockchain workflow.

### 2.6 Stochastic models

The word "stochastic" means "chance" or "random" [11]. The antonym of it is "certain", "sure", or "deterministic". A certain outcome can be obtained by a deterministic model and possible outcomes can be predicted by the use of stochastic models. Normally, standard modeling consists of three components: (1) the phenomenon you want to research, (2) the logical system you should build to understand the phenomenon, (3) the connections between the natural phenomenon and the logical system [37]-[39].

By assuming the random components of the model, stochastic models could capture the random variables and test the predictions by statistical analysis. The computation of stochastic models can be complicated and requires different mathematical methods. We classify the common stochastic models in blockchain into four types with reference to [15]: queueing model; Markov Process, e.g., Discrete Time Markov chain (DTMC) and Continuous Time Markov Chain (CTMC); Markov Decision Process (MDP); Hidden Markov process (HMM). The basic knowledge about stochastic models can be found in [40]. They have their own strengths and the practicability of them has been tested for many years.

In blockchain-based platforms, queueing theory can be applied to the transaction execution process and block generation process. Some fork situations and performance evaluation can be accomplished by Markov Processes. It is convenient to choose optimal strategy by MDP [39]. HMM is a good technique for prediction [41]. Besides, stochastic models can be widely used to combine with other techniques, such as machine learning. The detailed analysis will be discussed in the next chapters.

TABLE I. A SUMMARY OF STOCHASTIC MODELS FOR PERFORMANCE

| | | REF. | Platform | Stochastic model | Assessment method | Research objective | Main performance metrics |
|---|---|---|---|---|---|---|---|
| Consensus algorithm | Raft | [44] | Private blockchain | Markov process | derive steady-state probability | analyze the performance of the Raft consensus algorithm | split probability |
| | PBFT | [45] | Hyperledger fabric | Markov process | give numerical analyses | investigate whether PBFT is a performance bottleneck for large networks | mean time to consensus |
| | PBFT | [48] | Healthcare blockchain network | Markov process | give numerical analyses | analyze the performance of the PBFT consensus algorithm | mean network delay |
| | PoA | [51] | Consortium blockchain | MDP | give numerical analyses | analyze the performance and security of the PoA consensus algorithm | stale block rate |
| | DAG | [53] | Blockchain | Markov process | derive transient probability | analyze the consensus process of DAG | cumulative weight |
| Platform | Hyperledger Fabric | [55] | Hyperledger fabric | Markov process | derive steady-state probability | analyze the performance of Hyperledger Fabric | transaction endorsement failure, block timeout |
| | Hyperledger Fabric | [54] | Hyperledger fabric | Markov process | give numerical analyses | analyze the performance of Hyperledger Fabric | throughput utilization, mean queue length |
| Transaction | | [56] | Bitcoin blockchain | queueing model | derive formulas | analyze the impact of Bitcoin fee on transaction confirmation process | mean transaction confirmation time |
| | | [57] | Blockchain | queueing model | derive formulas | give mathematical assesment of blocks acceptance in blockchain | transaction confirmation time |
| | | [58] | Bitcoin blockchain | queueing model | give numerical analyses | analyze the delay of transactions expected to be confirmed | mean time between transactions |
| | | [59] | Blockchain | queueing model | derive formulas | give discrete-time analysis of the blockchain distributed ledger technology | transaction waiting time |
| Other studies | | [61] | B-RAN | queueing model | derive steady-state probability | evaluate the performance of blockchain radio access network | system latency |
| | | [62] | Blockchain | Markov process | derive steady-state and transient probability | understand the blockchain evolution and dynamics | block dissemination delay |
| | | [65] | Bitcoin | queueing model | derive formulas | give a comprehensive analysis of Bitcoin's blockchain distribution network | forking probability, network partition sizes |
| | | [67] | Bitcoin distributed network | queueing model | derive formulas | give a detailed analysis of Bitcoin's blockchain distribution network | block delivery time, forking probability |
| | | [68] | Blockchain | Markov process | derive formulas | discuss the traffic generated by the synchronization protocols | traffic, protocol execution duration |
| | | [69] | Blockchain | queueing model-based | give numerical analyses | understand the working and theoretical aspects of the blockchain | system throughput, mining time of each block |

3. STOCHASTIC MODELS FOR BLOCKCHAIN NETWORKS PERFORMANCE

When blockchain networks bring tremendous benefits, people worry about whether their performance would meet the needs of the mainstream IT systems. Currently, using stochastic models to evaluate the performance of blockchain networks mainly focuses on consensus, platform, and transaction. Besides the three mainstreams, the rest topics about blockchain networks performance will be discussed in Section 3.4. The summary of stochastic models for performance is outlined in TABLE I. The definitions in TABLE I will be given in the following detailed discussions.

*3.1 Consensus algorithms*

Blockchain consensus mechanism is designed to ensure the consistency and correctness of each transaction among all nodes. As demand changes, various consensus protocols emerge. Through the performance analysis of these consensus algorithms, we can understand different consensus algorithms better and choose the more appropriate one. Currently, the application of stochastic models in consensus algorithms mainly includes Raft [44], PBFT [45], PoA [51], and DAG [53]. There are other consensus algorithms like PoS and PoR (Proof of Reputation) [42], but there is no modeling for these consensus algorithms. Therefore, we do not talk about them in this section.

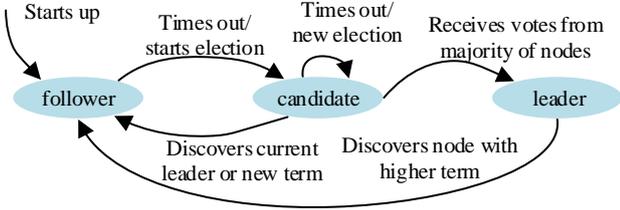

Fig. 5. State transition model for the Raft algorithm [44].

### 3.1.1 Raft

Raft achieves consistency by electing a leader and the leader is responsible for managing ledger replication. The state transitions of Raft are shown in Fig. 5. A more detailed description of Raft refers to [43].

In Raft, there is one leader and other nodes are followers. There is a situation called split, which means that more than half of the nodes are not controlled by current leader. Specifically, packet loss will lead to the failure of node and communication interruption, which are the main reasons for networks split. When the network split happens, a new leader election process will be restarted and the new transactions are rejected by the blockchain network at the same time, leading to unavailable blockchain network. For this problem, a Markov process (DTMC) is built to capture the process of Raft consensus algorithm [44]. Through the analytical model, many network performance metrics in normal conditions are derived, including network split and election time. The experiment results indicate that the larger the network, the smaller the split probability at the beginning of runtime. They also prove that the split probability caused by packet loss can be reduced by the increase of election timeout, but this conclusion does not hold in larger networks.

### 3.1.2 PBFT

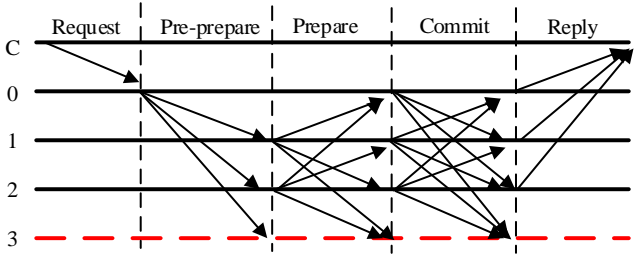

Fig. 6. The main process of PBFT.

PBFT process can be divided into several stages. The main process of PBFT (Practical Byzantine Fault Tolerance) is shown in Fig. 6.

As for PBFT, the authors in [45] aim to investigate whether PBFT consensus process would affect the performance of the networks with a large number of peers. The process of PBFT consensus is modeled as CTMC (Stochastic Reward Nets (SRN) [37]) to compute the mean time to complete consensus in a network of 100 peers and numerical analyses are given by SPNP tool [46]. To parameterize and validate their models, they use IBM Bluemix service [47] to create a blockchain network and run IoT application on it to obtain the data. The performance of larger networks is also examined. The experiment results show that (1) compared to the slowdown of preparing messages, the slowdown of handling incoming prepare and commit messages can have a greater impact on the mean time to consensus; and (2) the mean time to consensus does not increase significantly when the number of peers increases.

Different from [45], the authors in [48] use CTMC to model the process of PBFT in healthcare blockchain networks. The CTMC model is run by PRISM [49]. Based on the results of PRISM, mean network delay has a major influence on the finish time of the PBFT process. The impact of network delay among client and other nodes (primary and replica nodes) is analyzed. They conclude that network speed has the greatest impact on the probability of the whole process finish. However, the two papers about PBFT just discuss a small parts of performance aspects and a systematic analysis should be done.

### 3.1.3 PoA

Proof of Authority (PoA) is a novel Byzantine fault tolerant (BFT) algorithm. PoA aims at consortium blockchain that only the authorized nodes can join blockchain networks and submit transactions to the blockchain. PoA algorithms are different from BFT-like algorithms. PoA requires fewer message exchanges and can be deployed to a larger scale [50]. Moreover, in the presence of attackers, PoA could work regularly as long as the proportion of the adversaries is no more than half of the whole. Specifically, when there are $N$ nodes in the blockchain of PoA, the number of honest nodes is at least $N/2+1$.

However, limited analysis is conducted to prove the performance of PoA. The authors in [51] provide a quantitative analysis of PoA protocol. A MDP-based framework is presented to model the concrete PoA-based blockchain. An important metric called stale blocks rate is used to represent the performance of a blockchain system. Stale blocks refer to the blocks not linked to the main chain. Because stale blocks will cause forks, leading to slow chain growth. The more stale blocks, the more harm to the performance of blockchain systems. The authors use stale blocks rate to compare the performance of different PoA implementations. However, the framework they proposed is under the given parameters, which lacks a complete analysis in terms of formal proof.

### 3.1.4 DAG

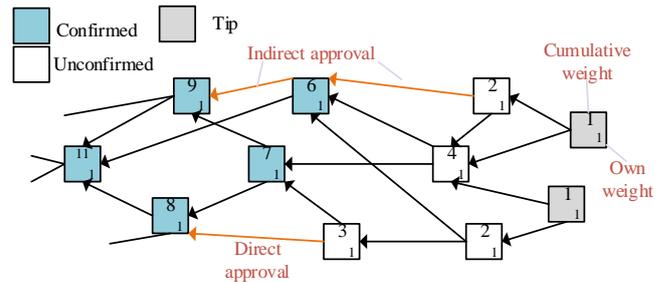

Fig. 7. An example of DAG consensus process.

Direct Acyclic Graph (DAG) consensus is firstly introduced in [52] and any node is allowed to insert transactions into blockchain networks. The transactions are organized in a topology of direct acyclic graph(see Fig. 7). The confirmation level of a transaction is represented by cumulative weight. When the cumulative weight reaches the defined threshold, the transaction can be confirmed. The consensus process of DAG can be found in [53].

Compared to PoW and PoS, DAG consensus has many advantages like fast transaction speed, no transaction fee and no mining required. These advantages of DAG based blockchain well meet the needs of IoT, but the theoretic analysis is not sufficient. A Markov chain is built to model the consensus process of DAG in dynamic load conditions [53]. The authors model the situations of high to low and low to high load regime with DTMC. They evaluate the impact of the network load on the cumulative weight and confirmation delay. The transient probability of cumulative weight is derived. Through extensive experiments, insightful results are presented: the confirmation time will be very long under the situation of network load changing from high to low. Instead, the transactions can be confirmed fast in the environment of network load changing from low to high. Unfortunately, the proposed mathematical models cannot be applied to other DAG based blockchain networks due to different characteristics among the consensus processes.

Some consensus algorithms can be well combined with stochastic models like Raft and PBFT. Their consensus process can be divided into steps. The nodes or the roles in consensus algorithms can be models as states in Markov chains and the transformations between steps can be modeled as transitions between states in Markov chains. In addition, to validate the accuracy of models, most works adopt simulation without actual experiment environments. Only [45], they build a real blockchain network to obtain data and further prove the correctness of their results. Another thing to note is that consensus protocols have application restrictions. For example, Raft is applicable to permissioned blockchain and PoA aims at consortium blockchain. When stochastic models are built, these conditions should be noticed.

*3.2 Blockchain platform*

Since Bitcoin, many blockchain platforms have emerged like Ethereum, Hyperledger Fabric and so on. Usually, Bitcoin and Ethereum are discussed together with other performance indicators, such as transactions in Bitcoin and forks in Ethereum. So we classify them in other sections. In this section, we mainly introduce Hyperledger Fabric. Hyperledger Fabric is one of the most popular blockchain platforms in permissioned blockchain [3]. There are three major stages in Hyperledger Fabric when a transaction is processed, namely, endorsement phase, ordering phase and validation phase (see Fig. 8):

1) Endorsement: Transactions are sent from a client to the peers according to the endorsement policy. The client verifies the consistency of the received results after it receives endorsement from all the involved peers.

2) Ordering: In the ordering phase, transactions are grouped into blocks, which could improve the throughput of the platform.

3) Validation: Once nodes receive a new block, transactions are validated to ensure that the endorsement policy is satisfied. Then, the block will be appended to the blockchain.

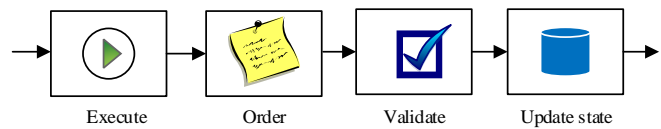

Fig. 8. Execute-order-validate architecture of Hyperledger Fabric.

In Hyperledger Fabric, the participants know and identify each other but do not fully trust each other. During the process of its evolution and development, it is important to model the interactions among peers performing different functions. Motivated by this, the authors in [54] provide a performance model of SRN to compute the mean queue length, utilization, throughput at various peers. There are one peer, one client, and two endorsing peers in the SRN model. And the model ensures that the corresponding sub-models are pluggable. Through the numerical analyses of SPNP, they find that the number of peers and policies have a large impact on the time to complete the endorsement process. However, since the committing peer validates transactions in parallel, their approach has a problem that the number of states in their model is very large.

To solve the problem of [54], Jiang et al. [55] develop a hierarchical model to analyze the performance of Hyperledger Fabric from the time when clients submit the transactions until the transactions are validated as a block. A monolithic model is introduced for each sub-process, namely, Transaction Execution (TE) and Transaction Validation (TV). The TE process is modeled as CTMC with *m* endorsing peer for providing the endorsing service. After the TE finishes, the transactions enter TV process that is in charge of batching and validating phases. The formulas for calculating various metrics are given. Besides, the accuracy of the model is verified by the simulation and numerical analyses. The results indicate when the transaction arrival rate is fixed, the mean response delay increases first and then decreases with the increase of block size. It is less good that all time intervals obey exponential distribution.

*3.3 Transaction*

Transaction is the records movement of cryptocurrency between users. Transaction fee is created by clients and will be given to the miner who mines successfully. There are two incentives for miners to mine Bitcoin blocks. After mining a block successfully, the mining reward of creating a new block and the transaction fees included in the block will be given to the miner. Generally, the mining reward is larger than the transaction fees. Only transactions in blocks included in the blockchain are admitted as valid ones, which are called confirmed transactions [57].

As for transaction, transaction confirmation time is an important performance metric. Bitcoin transactions usually take a short time (minutes) to confirm, but still larger than traditional credit card systems (seconds). VISA and Paypal could handle 450 and 1,667 transactions per second but Bitcoin and Ethereum can only process four and 20 transactions per second respectively. The slow transaction rate hampers the scalability of blockchain. Many researches have been done to discuss the influence of different factors on it.

Kasahara et al. [56] study the impact of transactions with small fees of bitcoin on the transaction confirmation time. A priority queueing system is created based on queueing theory to research the transaction confirmation time. Through the model, they derive the formula of the mean transaction confirmation time. By numerical analyses, they show that transactions with small fees suffer from an extremely large confirmation time.

Different from [56], the authors in [58] care about the delay incurred by confirming transactions. The delay in blockchain networks hinders the development of bitcoin. Motivated by that problem, a framework is proposed to identify which transactions will be confirmed and characterize the confirmation time. The framework combines machine learning with queueing theory. Machine learning is used to classify the transactions. Then, a queueing theoretic model is adopted to describe the delay of transactions. Their model considers the aspects related to transaction delays, like the mean time between transactions and the activity time of blocks. By numerical analyses, their paper concludes that the delay is slightly larger than the time between block generations.

Srivastava [57] assesses the transaction confirmation time and the acceptance rate of blocks by Markov model. Based on the model, the formula of the transaction confirmation time is given. The correctness of the model is verified by the experiment results.

Besides transaction confirmation time, the transaction waiting time is discussed in [59]. They study how different input parameters affect key performance features of blockchain. A discrete-time queueing model is developed to evaluate the features of a blockchain system. By the model, the waiting time distribution of transactions is displayed and they investigate the impact of different transaction size on the mean waiting time. The experiment results show that the mean waiting time will not increase all the time as transaction size increases.

The studies about transaction are mainly researched by queueing models. Transactions arrive at systems according to a specific rate and are stored in a queue. Although assessing different parameters and executed in different settings, the models of these papers are similar.

### 3.4 Other performance evaluation

In order to provide better services to mobile devices, blockchain radio access network (B-RAN) has emerged as a decentralized and reliable radio access example supported by blockchain technology. But the characteristics of B-RAN are still unclear and need to be analyzed. Ling et al. [61] use a time-homogeneous Markov chain [40] to establish a queueing model starting from block generation and evaluate the performance of blockchain radio access network. The steady-state probabilities of many metrics are derived, such as the average of number of waiting requests and average access latency. Afterward, they further prove that block arrival forms a Poisson process. Through their work, they establish an original framework to study the property and performance of B-RAN.

When blockchain is deployed to industry, the nodes in blockchain are distributed in different locations. The processing delay and block transfer will become key issues. Motivated by it, Papadis et al. [62] develop a stochastic model to analyze how hashing power of the nodes and block dissemination delay influence the performance of blockchain. The authors combine theoretical analysis with simulation experiment to investigate both stationary and transient performance features. To check their results, they show when the block arrival is not strictly homogeneous, their conclusions match the results in the Ethereum tested [63].

In [65], a comprehensive analytical model is provided to understand bitcoin's blockchain distribution network. The data distribution algorithm is modeled by branching process [66]. The individual nodes are modeled as priority M/G/1 queuing systems. Their model could analyze a number of performance metrics, including forking probability [64], node response time, and network distribution time. The probability distributions of many metrics are also given. Their results give strong qualitative and quantitative analyses of network performance. As for fork probability, their results show that block size and node connectivity are major factors that affect fork probability. Similarly, they use almost the same methods to analyze block delivery time in [67]. Their results show that intensity of transaction traffic has little effect on the performance of block traffic.

Besides the above introduction, there are other researchers studying performance about blockchain. Danzi et al. [68] model the traffic generated by the synchronization protocols as a Markov process. They investigate several protocols for synchronization about IoT and blockchain networks. The distributions of protocols execution time are given. They want to prove how the blockchain parameters and communication link quality influence the synchronization process. The correctness of the analytical model is verified by numerical analyses. The results show that if the execution duration of protocol is equal to the time of block-generation, the probability of keeping synchronized will decrease rapidly.

To understand the working and theoretical aspects of the blockchain, Memon et al. [69] introduce a queueing theory-based model. The memory pool is modeled as M/M/1 and the mining pool is model as M/M/c. The author first gets the performance metrics from the model and then compares them with actual data. There is a conclusion that the results obtained from the model match the actual statistical data well.

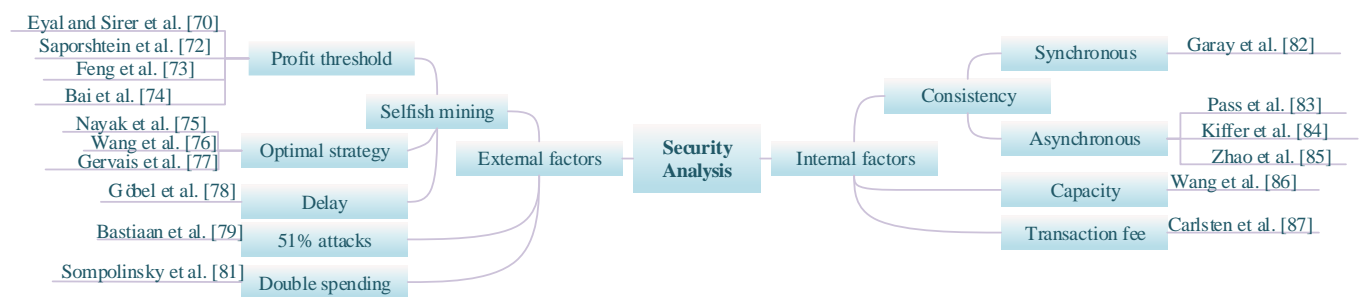

Fig.9. Application of stochastic models for security analysis.

## 4. STOCHASTIC MODELS FOR BLOCKCHAIN NETWORKS SECURITY

As presented in Fig.9, according to the causes of security problems, stochastic models for security analysis can be divided into two directions: (1) security problems coming from external factors, such as various attacks. (2) security problems coming from internal factors, such as the natural properties of blockchain networks. As for the privacy of blockchain, no work analyzes it by stochastic models, so it is not included in our survey. The detailed explanations are as follows.

### 4.1 Security problems coming from external factors

Among these external factors, malicious attacks are the main threats. Especially, selfish mining analyzed by stochastic models accounts for a large proportion

#### 4.1.1 Selfish mining

Selfish mining is proposed in [70]. In order to reach global consensus, the miners need to solve a Hash puzzle, which is known as PoW. When the miner is the first one to find a valid block, he is awarded by cryptocurrency (i.e. bitcoin). The blocks should be published immediately after they are mined. However, Eyal and Sirer present an attack and the miners could obtain a larger revenue than their fair share by obeying the strategy. The strategy develops some principles: (1) *Risk-avoiding release case*. When the leading advantage of selfish miner's private chain is no more than two blocks, the selfish miner releases his (her) mined blocks to the public because of the fear of loss. (2) *Tie-breaking resolving*. When the number of blocks of an attacker is equal to the honest miner, the attacker publishes his private blocks to compete with the honest miners. In other situations, the leading advantage of selfish miner's private chain is obvious, the selfish miners will keep the blocks secret in private chain.

Overall, the selfish miner could obtain more revenue by withholding blocks strategically. Since selfish mining is introduced, many research studies have been conducted based on it. Selfish mining is proposed under the platform of bitcoin. The authors in [71] extend the definition into Ethereum and calculate the uncle rewards. Many researchers show that selfish mining is not the optimal strategy and they are studying new strategies to get more rewards.

Recently, the attraction of selfish mining mainly concentrates on profit threshold and optimal strategy. There will be a detailed explanation next.

**(1) Profit threshold**

Markov process is one of the common techniques to capture the behaviors of selfish mining, which is originated from Eyal and Sirer. Among the research studies, the threshold of computational power that makes selfish mining profitable is an important metric. In 2014, Eyal and Sirer [70] first proposed the definition of selfish mining and they proved when the power controlled by selfish miner is no more than 25% of the resources, the selfish mining can be prohibited. In their model, the selfish mining becomes profitable only when miners possess enough computational resources, and connect with the network well.

However, in 2016, a more intelligent selfish miner in [72] can low down this threshold to around 23.21% by the use of related Markov Decision Process (MDP). An efficient algorithm is given to compute the optimal selfish mining policy and they prove the correctness of their algorithm. They show that compared to [70], there are other strategies that earn more rewards and are profitable for fewer miners. They also indicates that if the model considers the delay of block propagation in the network, the attackers of any scale would benefit from selfish mining.

Besides, the authors in [73] investigate the selfish mining in Ethereum. They build a 2-dimensional Markov process to capture the behavior of selfish mining and they find when the hash rate of selfish miner is above 16.3%, more revenue can be gained by the selfish pool from selfish mining. Stationary distribution is derived and long-term average mining reward is also computed too. The experiment shows when the hash rate of selfish miner is below the threshold 16.3%, the selfish pool loses just a small part of revenue. Because uncle block rewards are added in the computation of profit, which is different from Bitcoin.

In the above studies, the selfish miners are modeled as a selfish pool. In 2019, Bai et al. [74] study the reward of selfish mining when there are multiple selfish miners in blockchain networks. They assume that there are more than one selfish miners and one honest miner, but the maximum length of private chain is limited in the model. A CTMC is built to describe the evolution of private and public chains. The threshold of their model reduces to 21.48%. The comparison of the above models can be found in TABLE II. In general, Ethereum is more vulnerable to selfish mining than Bitcoin and the more selfish miners, the lower the threshold.

TABLE II. MODELS COMPARISON ABOUT PROFIT THRESHOLD

| REF. | Platform | Stochastic model | Profit threshold | The number of selfish miners |
|---|---|---|---|---|
| [70] | Bitcoin | CTMC | 25% | one |
| [72] | Bitcoin | MDP | 23.21% | one |
| [74] | Bitcoin | CTMC | 21.48% | more than one |
| [73] | Ethereum | CTMC | 16.3% | one |

**(2) Optimal strategy**

We all know selfish miners can get improper rewards under certain strategies. Recent researches [75]-[77] show that selfish mining attacks may not be optimal.

In [75], the authors design a new strategy called stubborn mining. The stubborn mining means that the attacker should not give up quickly even if the private chain will fall behind the public chain. The attacker still increases its profit by mining on its private chain more frequently (stubborn). The result shows that stubborn mining can result in 30% of gains with combination of eclipse attacks.

Although there are many versions of selfish mining, it is elusive to find the most profitable mining strategy until [72]. To obtain the optimal mining strategy, the mining problem is formulated as a general MDP in [72]. However, the objective function of the mining MDP is not linear, so a standard MDP solver cannot solve it. To address the problem, the mining MDP with the non-linear objective is transformed to a family of MDPs with linear objectives and then a standard MDP solver is utilized to find the optimal mining strategy. In the model of [72], we need to know various parameters before the MDP can be established. But it is not easy to get the exact parameter values in real blockchain networks. A model-free approach [76] is proposed to solve the problem of [72].

Reinforcement learning is employed to solve the mining MDP and the parameter values do not need to be known. They first adopt the same mining model as [72] and then present a reinforcement learning algorithm to reason about the optimal mining strategy without knowing the parameter values. In dynamic environments, the method of [76] can be more robust.

Different from the purpose of the papers above, Gervais et al. [77] provide a novel quantitative framework to analyze the security of PoW (see Fig. 10). Their framework includes two parts: a blockchain instance and a blockchain security model. The blockchain security model is based on MDP for selfish mining and could derive optimal adversarial strategy while considering the adversarial mining power. They show that selfish mining is not always a rational strategy. They also discuss the selfish mining and optimal double-spending strategies based on the security model. Following their optimal mining strategy, an adversary can yield 209 blocks rewards on average when the whole network mines 1000 blocks. While the strategy in [70] generates 205.8 blocks rewards on average.

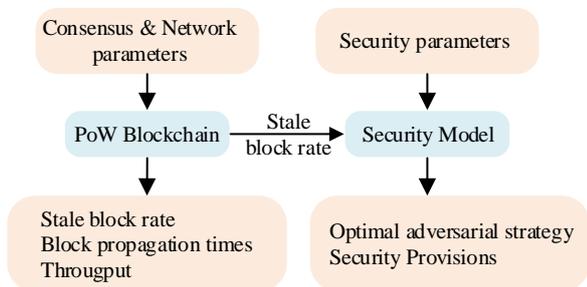

Fig. 10. Components of [77] quantitative framework.

Besides profit threshold and optimal strategy, Göbel et al. [78] study the selfish mining strategy in the existence of the propagation delay. They use a simple Markov model to track the state of dishonest miners that is different from the state of honest miners. Through their analysis, they prove that it is possible to detect block-hiding behavior in selfish mining by observing the rate of production of orphan blocks.

*4.1.2 Double spending and 51%-attacks*

Besides selfish mining, 51%-attacks [79] and double spending attacks [80] can also be analyzed by stochastic models. In [79], 51%-attacks are discussed in 2P-PoW. Two phase Proof of Work (2P-PoW) is a second cryptographic challenge, in which pool operators are forced to either give up their private keys or provide a substantial part of their pool's mining hash. This behavior would make pools smaller and prevent the encourage of large pools that is vulnerable to 51%-attacks. The author in [79] uses CTMC to model the Bitcoin mining protocol extended with 2P-PoW. By PRISM [49], the correctness of the model is verified by simulation. They prove that the benefit of 2P-PoW is good at preventing the 51%-attacks.

If a certain coin is spent twice or more times, and then this phenomenon is known as double spending. It is impressed that the blockchain may experience a reorganization when the attacker publishes a longer chain. Therefore, a transaction should not be confirmed because the attacker may replace it with another in a double spending attack. To overcome this problem, transactions are designed to be confirmed after five blocks following the block that contains the transaction. However, the paper [80] puts forward a new idea that the success of double spending attacks not only depends on the number of confirmations but the time elapsed since transaction is broadcast. To derive the expression for the probability of double spending, the authors in [81] study the idea of [80]. A CTMC is presented, demonstrating the process of double spending shown in Fig. 11. The numbered states mean the number of blocks that the attacker is ahead. The labels between the states stand for transition rates. The probability of double spending can be obtained in the Markov process and formula of the probability of double spend is introduced. By numerical results, they find when the expected time required for transaction acceptance improves, the time and cost to perform a double spend attack reduce.

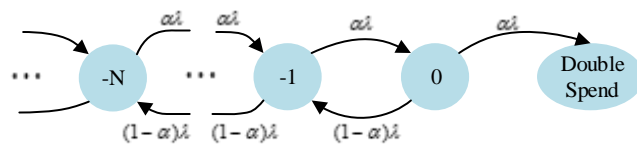

Fig. 11. Markov process model of double spending [89].

*4.2 Security problems coming from internal factors*

In addition to external attacks, the natural properties of blockchain networks can also become threats. Consistency is a guarantee that all honest parties output the same sequence of blocks by executing the protocol. Consistency is a security property of blockchain, so we discuss it in this section.

The celebrated Nakamoto proposes a PoW protocol to achieve a public, immutable and ordered ledger of records, which could function in a permissionless setting (anyone could join or leave) and work as long as more than 50% computing power follows the protocol. However, Byzantine agreement shows that as long as the proportion of malicious nodes is less than one third, the consistency can be achieved. Recently, there are studies showing a lower bound to accomplish consistency. The research works can be divided into two directions, synchronous setting (sync) and asynchronous setting (async).

**(1) Synchronous**

In synchronous settings, the authors in [82] prove that the blockchain protocol satisfies consistency by analyzing the property of common prefix and chain quality. And they provide the first formal model of Nakamoto consensus but they prove consistency under high synchronization and the number of players remains fixed, which is not suitable with the actual situation. The assumption of synchronization is strong, and actually, Nakamoto protocol is designed in a network with message delay.

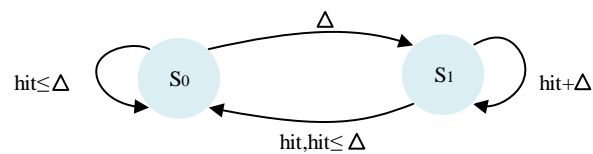

Fig. 12. A simple Markov model [84].

**(2) Asynchronous**

In asynchronous settings, a stronger notion of consistency as T-consistency [83] is introduced. T-consistency means that except for T unconfirmed blocks at the end of the chain, honest players reach consensus on the current chain. They observe, in the case of complete asynchrony where an adversary can arbitrarily delay messages, there is no guarantee of consistency. Therefore, their work is conducted in the setting of partial synchrony (that is, the adversary could delay messages at will as long as it passes them in time $\Delta$). They prove Nakamoto could achieve T-consistency as long as the hardness parameter exceeds $1/\rho n\Delta$ where $\rho$ is the proportion of the computational power occupied by the attacker and there are $n$ players in the network. Besides, the research [83] is the first one to consider the spawning of new players.

Relying on the research of [82]-[83], a simple framework of Markov processes (seen in Fig. 12) is introduced [84] to model blockchain protocols. Markov-based models are used to analyze consistency firstly under a partially asynchronous setting. By the analysis of their model, the consistency of blockchain protocols is further guaranteed. Their model replicates the analysis of Nakamoto's protocol done by [83] and obtains the same bound as it, which verifies the correctness of their method. They also make use of their method to analyze three other blockchain protocols such as GHOST, Nakamoto, and Cliquechain. Their techniques can also be applied to analyze the properties about subsequent protocols.

Inspired by [84], Zhao et al. [85] formulate a novel Markov chain in an asynchronous network to derive a neat bound by $c > 2\mu/\ln(\mu/v)$ where $c$ means the expected number of network delays before the block is mined and $\mu$ (resp., $v$) denotes the fraction of computational power held by honest (resp., the attacker) miners. Their conclusion is stronger than the results of existing ones. However, they [84]-[85] do not deal with the change of players in blockchain systems. The following TABLE III gives the comparison of the above studies and relevant parameters can be found in [82] or [83].

TABLE III. MODELS COMPARISON ABOUT CONSISTENCY

| REF. | Sync | Async | Proved properties | New players | Bound |
|---|---|---|---|---|---|
| [1] | | √ | no proved | Yes | 1/2 |
| [36] | | √ | no proved | No | 1/3 |
| [82] | √ | | common prefix, chain quality | No | $\lambda^2 - p\lambda + 1 \geq 0$ |
| [83] | | partially | consistency, liveness | Yes | $p > \dfrac{1}{\rho n\Delta}$ |
| [84] | | partially | consistency | No | same as [83] |
| [85] | | √ | consistency | No | lower than [83] |

Apart from consistency, there are other internal factors could affect the security of blockchain. For example, blockchain capacity is an important consideration in IoT. Devices in IoT are immense scale and resource-limited. Data integrity is vulnerable in IoT (e.g., to tampering). As a distributed ledger database, blockchain has the potential to solve the security issues of IoT. The authors in [86] build a testbed on the Ethereum platform and propose a DTMC to validate their testbed. Both the testbed and analysis show that accelerating the block mining rates could improve the blockchain capacity, but the disadvantage is that this approach will increase stale blocks at the same time.

The impact of transaction fees on bitcoin mining strategies is researched in [87]. It is said that the mining fee is designed to diminish in the future and the transaction fee will take its place. A CTMC is adopted to study how transaction fees affect security in the bitcoin network. The authors prove that this change can lead to uneven blocks and will increase the opportunity of forking. They reexamine the selfish mining strategy, which shows that in a transaction fee environment, any miner could get profit regardless of their hash power.

5. STOCHASTIC MODELS FOR BLOCKCHAIN APPLICATIONS

Besides performance and security in blockchain networks, the widest adoption of stochastic models in blockchain is the prediction of cryptocurrency price. Usually, the stochastic models would combine with machine learning to predict the trend of cryptocurrency.

It is well-known that cryptocurrency is untraceable and uncontrolled and it will be a challenge to predict the price of it. In recent years, many people have tried to solve the problem according to the cryptocurrency's past price. Normally, with previous data, machine learning is the most commonly used technique to predict the price of cryptocurrency [88]-[90]. The most important task of machine learning is to infer unknown variables of interest (such as class labels) based on observed data (such as training samples). Hidden Markov Model (HMM) is an important probability model and able to integrate well with machine learning, which has been widely used in speech recognition [91], natural language processing [92] and fault diagnosis [93]. According to the used methods, the related works in blockchain application can be divided into two categories, namely, HMM and MSGARCH. In previous research, HMM has been used to predict the finance fluctuation [94] and with the development of blockchain, HMM changes into a tool to predict the cryptocurrency price.

*5.1 HMM*

Based on cryptocurrency's historical movements, the authors in [95] introduce a hybrid model for bitcoin price prediction. They describe the cryptocurrency's historical movement with HMM and predict future movements by Long Short Term Memory networks. Their model emphasizes the dynamics of bitcoin prices by capturing hidden information. Koki et al. [96] analyze both ether and bitcoin prices by the Non-Homogeneous Pólya Gamma Hidden Markov model (NHPG) [97] that performs well in non-linear and linear effects of the predictors. They unexpectedly find that although the prices of bitcoin and Ethernet are related, they are greatly influenced by different variables. Phillips et al. [98] aim to predict the bubbles [99] by HMM. In [100], they research the relationship among the social metrics, coin market behavior, and market events. They examine the situation of the coin market. They also present that market events like trading volume and closing price could predict the coin market behavior. And HMM shows that the future behavior of the coin markets is explained in terms of social media metrics. Markov regime-switching (MRS), also called HMM, is often used to explain regime heteroscedasticity in the returns of financial assets. Daniel Chappell [101] aims to identify the optimal number of states for modeling the regime heteroscedasticity. By applying MRS, they find the evidence of shock persistence, volatility clustering, volatility jumps, and asymmetric volatility transitions.

*5.2 MSGARCH*

High and erratic volatility is an important characteristic of cryptocurrencies. There are many methods describing this behavior, and among them, Markov switching generalized autoregressive conditional heteroscedasticity (MSGARCH) [102] models are often used to solve the weaknesses of conventional standard deviation estimates. Using MSGARCH, [103]-[104] aim to understand the volatile behavior in different states. The results of [103] help us to understand the impact digital currencies on volatility states both in relation to negative and positive changes in the price of the cryptocurrency. While [104] shows to us that bitcoin's conditional volatility displays two regimes: high volatility and low volatility. TABLE IV presents the models comparison about cryptocurrency price prediction.

TABLE IV. MODELS COMPARISON ABOUT CRYPTOCURRENCY PRICE PREDICTION

| REF. | Stochastic model | Research objective | Currency type |
|---|---|---|---|
| [95] | HMM | describe cryptocurrencies historical movements | bitcoin price |
| [96] | NHPG | identify linear and non-linear effects of the predictors. | bitcoin and Ether price |
| [98] | HMM | predicte bubbles in cryptocurrency | cryptocurrecy |
| [100] | HMM | examine the interactions between the social metrics, coin market bahavior and market events | cryptocurrency |
| [101] | HMM | identify the optimal number of states for modelling the regime heteroscedasticity | bitcoin |
| [103] | MSGARCH | understand the relation between volatility and digital currency | cryptocurrency |
| [104] | MSGARCH | show the Bitcoin's conditional volatility | bitcoin |

## 6. CHALLENGES AND FUTURE DIRECTIONS

In the above sections, we present a survey on stochastic models in blockchain. Although blockchain has been greatly developed, some open issues still exist. Based on the above analysis, we now discuss and highlight the challenges for combining stochastic models and blockchain technology in this section. Some future directions are also presented.

As mentioned earlier, stochastic models have some weaknesses. The blockchain is a kind of decentralized system, not all components of which can be described by stochastic models. Therefore, in the process of analyzing the characteristics of blockchain, some indicators are ignored or the assumptions do not accord with the actual situations. For example, the authors in [81] focus on the behavior of attackers in double spending but without considering the propagation delay and block size. Although they derive the probability of double spending, the outcome of them is not convinced to people and cannot be applied in practice for the reason that the blockchain network is designed to operate in a permissionless setting and with propagation delay. To address the challenge, the stochastic models can be used for the scenarios that take into account as many indicators as possible.

*6.1 Performance limitation and possible directions*

As discussed above, due to its inherent distributed and point-to-point nature, compared with the traditional database system, the performance of blockchain needs to be improved. Currently, the throughput of blockchain platforms is several orders of magnitude lower than that of traditional database systems. If we want to apply blockchain to actual production environments, the throughput of blockchain networks must be improved. Apart from throughput, transactions in blockchain also need to be developed. It takes tens of minutes to complete the transaction on blockchain platforms, while it only takes seconds to complete the transaction in VISA. Therefore, it is a challenge to reduce the processing latency to a few minutes and maintain security at the same time. Moreover, because each node keeps a copy of ledger, data exchange in the network consumes the network bandwidth. With the increase of the scale of blockchain system, it is necessary to solve the bottleneck of network bandwidth.

For future directions, we can see there are some specific consensus protocols that are proposed to solve the problem of specific blockchains. For example, the Raft consensus protocol has the advantage of the high speed of transaction confirmation, so it is more applicable in permissioned blockchain. Therefore, according to certain situations, combined with actual scenes such as healthcare, Internet of Things and insurance, the research of consensus algorithms can be a good prospect. In addition, the performance of blockchain can be studied in many aspects, such as forking probability, transactions, block, and chains to be attached. Stochastic models can be used for analyzing more comprehensive scenarios from creation to confirmation of a transaction.

Another obvious issue about performance analysis is that some models that are built earlier should be updated according to the new research findings. In former papers, block arrival process is modeled as a homogeneous Poisson process [64] and several studies are conducted based on the assumption. However, the recent studies [62] show the process of block arrival is influenced by both difficulty adjustment and dynamic hash rate. Based on the new finding, the process of block arrival is proved to be a nonhomogeneous Poisson process. Before this conclusion is found, some papers have evaluated the performance of blockchain on the basis of previous researches. When the new study appears, some previous conclusions should be updated to discover new results, which should be worth our attention.

*6.2 Security limitation and possible directions*

Recent studies [83]-[85] are trying to prove consistency by building formal models in fully asynchronous settings, but every research has some shortcomings and they either do not prove consistency in a completely asynchronous setting or ignore the change of players. In the future, there is a need to consider both of the two conditions. The authors in [85] use Markov chain to characterize the numbers of mined blocks and prove the correctness of the neat bound. However, the results illustrate that it is stronger than others in consistency but weak in modeling the attacks. In the future, analyzing consistency and considering attacks at the same time need to be further developed in a (partially or fully) asynchronous setting. Also, stochastic modeling can be a favorable tool to understand consistency.

Selfish mining is an attack that can make attackers obtain more profits, which motivates people to devise new strategies. We believe, in the specific setting, such as IoT, healthcare, and industry, there must be better strategies. Analytical models can also be a good method to study the strategies in specific scenes. From TABLE II, we can know that studies have concentrated on Ethereum, but all the papers adopted the longest chain principles. Actually, Ethereum is designed by Greedy Heaviest-Observed Sub-Tree (GHOST) and there is no model to study selfish mining according to GHOST principles, which can be a future research direction. In addition, the previous studies have assumed the existence of a single honest miner and have not paid much attention to the situation where there were more than one honest miners in the system. When there are multiple honest miners, there may be forks. It is not sure when the forks will happen, which is difficult to model the behavior of several honest miners. As for the profit threshold, it shows a downward trend. It is expected that there is a lower bound for further research.

### 6.3 Cryptocurrency price prediction limitation and possible directions

After the emergence of machine learning, HMM is often combined with machine learning to predict something of interest based on the previous data. Nowadays, blockchain provides us with enormous data. Exploiting the data of blockchain, there must be other scenes that need further research. In addition, there are many methods, such as MSGARCH that are originally used in the traditional financial industry to evaluate the interaction among these currencies. Nowadays, the previous typical analysis methods can also be employed to solve issues in blockchain. Some practical models and techniques need us to explore and apply them in blockchain.

In the future, it is not limited to the cryptocurrency price prediction and stochastic models should be extended to the whole market. What is more, new prediction models need to be explored based on blockchain.

### 7. CONCLUSIONS

In this paper, we summarize the researches which use stochastic models for analyzing blockchain. Firstly, we introduce the fundamentals of blockchain and stochastic models. Then, we mainly classify the related works into two categories: network-oriented and application-oriented. We also have provided reviews and analyses in detail to deal with various problems regarding the two categories. Finally, We give challenges and suggestions for future studies. Through the above analyses, we hope our work could provide some references for others.

### REFERENCES

[1] Satoshi Nakamoto, Jungho Bang, Shengtuo Hu: A peer-to-peer electronic cash system[J]. Bitcoin.–URL: https://bitcoin. org/bitcoin. pdf, 2008.

[2] Dejan Vujičić, Dijana Jagodić, Siniša Ranđić: Blockchain technology, bitcoin, and Ethereum: A brief overview[C]//2018 17th international symposium infoteh-jahorina (infoteh). IEEE, 2018: 1-6.

[3] Elli Androulaki, Artem Barger, Vita Bortnikov, Christian Cachin, Konstantinos Christidis, Angelo De Caro, David Enyeart, Christopher Ferris, Gennady Laventman, Yacov Manevich, Srinivasan Muralidharan, Chet Murthy, Binh Nguyen, Manish Sethi, Gari Singh, Keith Smith, Alessandro Sorniotti, Chrysoula Stathakopoulou, Marko Vukolic, Sharon Weed Cocco, Jason Yellick:Hyperledger fabric: a distributed operating system for permissioned blockchains. EuroSys 2018: 30:1-30:15.

[4] Tianyi Qiu, Ruidong Zhang, Yuan Gao: Ripple vs. SWIFT: Transforming Cross Border Remittance Using Blockchain Technology. IIKI 2018: 428-434.

[5] Oscar Novo: Blockchain Meets IoT: An Architecture for Scalable Access Management in IoT. IEEE Internet of Things Journal 5(2): 1184-1195 (2018).

[6] Kamanashis Biswas, Vallipuram Muthukkumarasamy: Securing Smart Cities Using Blockchain Technology. HPCC/SmartCity/DSS 2016: 1392-1393.

[7] Mehdi Benchoufi, Philippe Ravaud: Blockchain technology for improving clinical research quality[J]. Trials, 2017, 18(1): 335.

[8] Yogita Khatri. 2018. Nearly $1 Billion Stolen In Crypto Hacks So Far This Year: Research. https://www.coindesk.com/nearly-1-billion-stolen-in-crypto-hacks-so-far-this-year-research.

[9] Today Altcoin: Bitcoin and Ethereum vs Visa and PayPal–Transactions per second-Altcoin Today[J]. Altcoin Today crypto news, 2018.

[10] Vermeulen Jan: Bitcoin and Ethereum vs Visa and PayPal-Transactions per second[J]. My Broadband, 2017, 22.

[11] Howard M.Taylor, Samuel Karlin: An introduction to stochastic modeling[M]. Academic press, 1998..

[12] Ying Liu, Kai Zheng, Paul Craif, Yuexuan Li, Yangkai Luo, Xin Huang: Evaluating the Reliability of Blockchain Based Internet of Things Applications[C]//2018 1st IEEE International Conference on Hot Information-Centric Networking (HotICN). IEEE, 2018: 230-231.

[13] Harish Sukhwani, Nan Wang, Kishor S. Trivedi, Andy Rindos: Performance Modeling of Hyperledger Fabric (Permissioned Blockchain Network). NCA 2018: 1-8.

[14] Tatsuo Mitani, Akira Otsuka: Traceability in Permissioned Blockchain. IEEE Access 8: 21573-21588 (2020).

[15] Wenbo Wang, Dinh Thai Hoang, Peizhao Hu, Zehui Xiong, Dusit Niyato, Ping Wang, Yonggang Wen, Dong In Kim: A Survey on Consensus Mechanisms and Mining Strategy Management in Blockchain Networks. IEEE Access 7: 22328-22370 (2019).

[16] Qi Feng, Debiao He, Sherali Zeadally, Muhammad Khurram Khan, Neeraj Kumar: A survey on privacy protection in blockchain system. J. Network and Computer Applications 126: 45-58 (2019).

[17] Shuai Wang, Liwei Ouyang, Yong Yuan, Xiaochun Ni, Xuan Han, Fei-Yue Wang:Blockchain-Enabled Smart Contracts: Architecture, Applications, and Future Trends. IEEE Trans. Syst. Man Cybern. Syst. 49(11): 2266-2277 (2019).

[18] Hong-Ning Dai, Zibin Zheng, Yan Zhang: Blockchain for Internet of Things: A Survey. IEEE Internet of Things Journal 6(5): 8076-8094 (2019).

[19] Jameela Al-Jaroodi, Nader Mohamed: Blockchain in Industries: A Survey. IEEE Access 7: 36500-36515 (2019).

[20] HaraldVranken: Sustainability of bitcoin and blockchains[J]. Current opinion in environmental sustainability, 2017, 28: 1-9.

[21] Seyoung Huh, Sangrae Cho, Soohyung Kim: Managing IoT devices using blockchain platform[C]//2017 19th international conference on advanced communication technology (ICACT). IEEE, 2017: 464-467.

[22] Karim Sultan, Umar Ruhi, Rubina Lakhani: Conceptualizing blockchains: characteristics & applications[J]. arXiv preprint arXiv:1806.03693, 2018.

[23] Martin Valenta, Philipp Sandner: Comparison of ethereum, hyperledger fabric and corda[J]. no. June, 2017: 1-8.

[24] Arati Baliga, Nitesh Solanki, Shubham Verekar, Amol Pednekar, Pandurang Kamat, Siddhartha Chatterjee: Performance Characterization of Hyperledger Fabric. CVCBT 2018: 65-74.

[25] Kenji Saito, Hiroyuki Yamada: What's so different about blockchain?—Blockchain is a probabilistic state machine[C]//2016 IEEE 36th International Conference on Distributed Computing Systems Workshops (ICDCSW). IEEE, 2016: 168-175.

[26] Deepak Puthal, Nisha Malik, Saraju P. Mohanty, Elias Kougianos, Gautam Das: Everything You Wanted to Know About the Blockchain: Its Promise, Components, Processes, and Problems. IEEE Consumer Electronics Magazine 7(4): 6-14 (2018).

[27] Florian Tschorsch, Björn Scheuermann: Bitcoin and beyond: A technical survey on decentralized digital currencies[J]. IEEE Communications Surveys & Tutorials, 2016, 18(3): 2084-2123.

[28] Ralph C. Merkle: A Digital Signature Based on a Conventional Encryption Function. CRYPTO 1987: 369-378.

[29] Sachchidanand Singh, Nirmala Singh: Future of financial and cyber security[C]//2016 2nd international conference on contemporary computing and informatics (IC3I). IEEE, 2016: 463-467.

[30] Mingxiao Du, Xiaofeng Ma, Zhe Zhang, Xiangwei Wang, Qijun Chen: A review on consensus algorithm of blockchain. SMC 2017: 2567-2572.

[31] Sunny Pahlajani, Avinash Kshirsagar, Vinod Pachghare: Survey on private blockchain consensus algorithms[C]//2019 1st International Conference on Innovations in Information and Communication Technology (ICIICT). IEEE, 2019: 1-6.

[32] JP Buntinx: What is proof of elapsed time?[J]. The Merkle Hash. Available online: https://themerkle. com/what-is-proof-of-elapsed-time/(accessed on 5 December 2019), 2017.

[33] Sarwar Sayeed, Hector Marco-Gisbert: Assessing blockchain consensus and security mechanisms against the 51% attack[J]. Applied Sciences, 2019, 9(9): 1788.

[34] Sunny King, Scott Nadal: Ppcoin: Peer-to-peer crypto-currency with proof-of-stake[J]. self-published paper, August, 2012, 19.

[35] Giang-Truong Nguyen, Kyungbaek Kim: A Survey about Consensus Algorithms Used in Blockchain. JIPS 14(1): 101-128 (2018).

[36] Miguel Castro and Barbara Liskov: Practical Byzantine fault tolerance[C]//OSDI. 1999, 99(1999): 173-186.

[37] Khalid Al-Begain, Dieter Fiems, Vincent Jean-Marc: Analytical and stochastic modelling techniques[J]. Annals of Operations Research, 2016, 239(2): 355-357.

[38] Vladimir S.Korolyuk, Vladimir V. Korolyuk: Stochastic models of systems[M]. Springer Science & Business Media, 2012.

[39] Wenlong Ni, Yuhong Zhang, Wei Wayne Li: Optimal Admission Control For Secondary Users using Blockchain Technology In Cognitive Radio Networks. ICDCS 2019: 1518-1526.

[40] Kishor S. Trivedi, Andrea Bobbio: Reliability and availability engineering: modeling, analysis, and applications[M]. Cambridge University Press, 2017.

[41] Roberto Casado-Vara: Stochastic Approach for Prediction of WSN Accuracy Degradation with Blockchain Technology. DCAI (Special Sessions) 2018: 422-425.

[42] Fangyu Gai, Baosheng Wang, Wenping Deng, Wei Peng: Proof of Reputation: A Reputation-Based Consensus Protocol for Peer-to-Peer Network. DASFAA (2) 2018: 666-681.

[43] Diego Ongaro, John Ousterhout: In search of an understandable consensus algorithm. in Proc. USENIX Conf. USENIX Annu. Tech. Conf.,Philadelphia, PA, USA, Oct. 2013, pp. 305–320.

[44] Dongyan Huang, Xiaoli Ma, Shengli Zhang: Performance Analysis of the Raft Consensus Algorithm for Private Blockchains. IEEE Trans. Systems, Man, and Cybernetics: Systems 50(1): 172-181 (2020).

[45] Harish Sukhwani, José M. Martínez, Xiaolin Chang, Kishor S. Trivedi, Andy Rindos: Performance Modeling of PBFT Consensus Process for Permissioned Blockchain Network (Hyperledger Fabric). SRDS 2017: 253-255.

[46] Gianfranco Ciardo, Jogesh K. Muppala, Kishor S. Trivedi: SPNP: Stochastic Petri Net Package. PNPM 1989: 142-151.

[47] "IBM Watson IoT Track and Trace contract - GitHub," https://github.com/ibm-watson-iot/blockchain-samples/.

[48] Kai Zheng, Ying Liu, Chuanyu Dai, Yanli Duan, Xin Huang: Model checking PBFT consensus mechanism in healthcare blockchain network[C]//2018 9th International Conference on Information Technology in Medicine and Education (ITME). IEEE, 2018: 877-881.

[49] Marta Z. Kwiatkowska, Gethin Norman, David Parker: PRISM: Probabilistic Symbolic Model Checker. Computer Performance Evaluation / TOOLS 2002: 200-204.

[50] Stefano De Angelis, Leonardo Aniello, Roberto Baldoni, Federico Lombardi, Andrea Margheri, Vladimiro Sassone: PBFT vs Proof-of-Authority: Applying the CAP Theorem to Permissioned Blockchain. ITASEC 2018.

[51] Xuefeng Liu, Gansen Zhao, Xinming Wang, Yixing Lin, Ziheng Zhou, Hua Tang, Bingchuan Chen: MDP-Based Quantitative Analysis Framework for Proof of Authority. CyberC 2019: 227-236.

[52] S. D. Lerner, "DagCoin draft," 2015. [Online]. Available: https://bitslog.files.wordpress.com/2015/09/dagcoin-v41.pdf.

[53] Yixin Li, Bin Cao, Mugen Peng, Long Zhang, Lei Zhang, Daquan Feng, Jihong Yu:Direct Acyclic Graph based Blockchain for Internet of Things: Performance and Security Analysis. CoRR abs/1905.10925 (2019).

[54] Harish Sukhwani, Nan Wang, Kishor S. Trivedi, Andy Rindos: Performance Modeling of Hyperledger Fabric (Permissioned Blockchain Network). NCA 2018: 1-8.

[55] Lili Jiang, Xiaolin Chang, Yuhang Liu, Jelena V. Misic, Vojislav B. Misic: Performance analysis of Hyperledger Fabric platform: A hierarchical model approach[J]. Peer-to-Peer Networking and Applications, 2020: 1-12.

[56] Shoji Kasahara, Jun Kawahara: Effect of bitcoin fee on transaction–confirmation process. arXiv preprint arXiv :1604.00103 .2016.

[57] Riktesh Srivastava: Mathematical assessment of blocks acceptance in blockchain using Markov model[J]. International Journal of Blockchains and Cryptocurrencies, 2019, 1(1): 42-53.

[58] Saulo Ricci, Eduardo Ferreira, Daniel Sadoc Menasché, Artur Ziviani, José Eduardo de Souza, Alex Borges Vieira: Learning Blockchain Delays: A Queueing Theory Approach. SIGMETRICS Performance Evaluation Review 46(3): 122-125 (2018).

[59] Stefan Geissler, Thomas Prantl, Stanislav Lange, Florian Wamser, Tobias Hoßfeld: Discrete-Time Analysis of the Blockchain Distributed Ledger Technology. ITC 2019: 130-137.

[60] Siamak Solat, Maria Gradinariu Potop-Butucaru: Zeroblock: Preventing selfish mining in bitcoin[J]. arXiv preprint arXiv:1605.02435, 2016.

[61] Xintong Ling, Yuwei Le, Jiaheng Wang, Zhi Ding, Xiqi Gao: Practical Modeling and Analysis of Blockchain Radio Access Network. CoRR abs/1911.12537 (2019).

[62] Nikolaos Papadis, Sem Borst, Anwar Walid, Mohamed Grissa, Leandros Tassiulas: Stochastic models and wide-area network measurements for blockchain design and analysis[C]//IEEE INFOCOM 2018-IEEE Conference on Computer Communications. IEEE, 2018: 2546-2554.

[63] "Geth", https://github.com/ethereum/go-ethereum/, Accessed: 07/27/2017.

[64] Vojislav B. Misic, Jelena V. Misic, Xiaolin Chang: On Forks and Fork Characteristics in a Bitcoin-Like Distribution Network. Blockchain 2019: 212-219.

[65] Vojislav B. Misic, Jelena V. Misic, Xiaolin Chang: Modeling of Bitcoin's blockchain delivery network[J]. IEEE Transactions on Network Science and Engineering, 2019.

[66] Geoffrey R. Grimmett, David R. Stirzaker: Probability and random processes[M]. Oxford university press, 2001.

[67] Jelena V. Misic, Vojislav B. Misic, Xiaolin Chang, Saeideh G. Motlagh, M. Zulfiker Ali: Block Delivery Time in Bitcoin Distribution Network. ICC 2019: 1-7.

[68] Pietro Danzi, Anders Ellersgaard Kalør, Cedomir Stefanovic, Petar Popovski: Analysis of the Communication Traffic for Blockchain Synchronization of IoT Devices. ICC 2018: 1-7.

[69] Raheel Ahmed Memon: Simulation model for blockchain systems using queuing theory[J]. Electronics, 2019, 8(2): 234.

[70] Ittay Eyal, Emin Gün Sirer: Majority is not enough: bitcoin mining is vulnerable. Commun. ACM 61(7): 95-102 (2018).

[71] Fabian Ritz, Alf Zugenmaier: The Impact of Uncle Rewards on Selfish Mining in Ethereum. EuroS&P Workshops 2018: 50-57.

[72] Ayelet Sapirshtein, Yonatan Sompolinsky, Aviv Zohar: Optimal Selfish Mining Strategies in Bitcoin. Financial Cryptography 2016: 515-532.

[73] Chen Feng, Jianyu Niu: Selfish Mining in Ethereum. ICDCS 2019: 1306-1316.

[74] Qianlan Bai, Xinyan Zhou, Xing Wang, Yuedong Xu, Xin Wang, Qingsheng Kong: A Deep Dive Into Blockchain Selfish Mining. ICC 2019: 1-6.

[75] Kartik Nayak, Srijan Kumar, Andrew Miller, Elaine Shi: Stubborn Mining: Generalizing Selfish Mining and Combining with an Eclipse Attack. EuroS&P 2016: 305-320.

[76] Taotao Wang, Soung Chang Liew, Shengli Zhang: When Blockchain Meets AI: Optimal Mining Strategy Achieved By Machine Learning. CoRR abs/1911.12942 (2019).

[77] Arthur Gervais, Ghassan O. Karame, Karl Wüst, Vasileios Glykantzis, Hubert Ritzdorf, Srdjan Capkun: On the Security and Performance of Proof of Work Blockchains. ACM Conference on Computer and Communications Security 2016: 3-16.

[78] Johannes Göbel, Holger Paul Keeler, Anthony E. Krzesinski, Peter G. Taylor: Bitcoin blockchain dynamics: The selfish-mine strategy in the presence of propagation delay. Perform. Eval. 104: 23-41 (2016).

[79] Martijn Bastiaan: Preventing the 51%-attack: a stochastic analysis of two phase proof of work in bitcoin[C]//Availab le at http://referaat. cs. utwente. nl/conference/22/paper/7473/preventingthe-51-attack-a-stochasticanalysis-oftwo-phase-proof-of-work-in-bitcoin. pdf. 2015.

[80] Yonatan Sompolinsky, Aviv Zohar: Secure High-Rate Transaction Processing in Bitcoin. Financial Cryptography 2015: 507-527.

[81] Seb Neumayer, Mayank Varia, Ittay Eyal: An Analysis of Acceptance Policies For Blockchain Transactions. IACR Cryptology ePrint Archive 2018: 40 (2018).

[82] Juan A. Garay, Aggelos Kiayias, Nikos Leonardos: The Bitcoin Backbone Protocol: Analysis and Applications. EUROCRYPT (2) 2015: 281-310 Pass R, Seeman L, Shelat A. Analysis of the blockchain protocol in asynchronous networks[C]//Annual International Conference on the Theory and Applications of Cryptographic Techniques. Springer, Cham, 2017: 643-673.

[83] Rafael Pass, Lior Seeman, Abhi Shelat: Analysis of the Blockchain Protocol in Asynchronous Networks. EUROCRYPT (2) 2017: 643-673.

[84] Lucianna Kiffer, Rajmohan Rajaraman, Abhi Shelat: A Better Method to Analyze Blockchain Consistency. ACM Conference on Computer and Communications Security 2018: 729-744.

[85] Jun Zhao: An Analysis of Blockchain Consistency in Asynchronous Networks: Deriving a Neat Bound. CoRR abs/1909.06587 (2019).

[86] Xu Wang, Guangsheng Yu, Xuan Zha, Wei Ni, Ren Ping Liu, Y. Jay Guo, Kangfeng Zheng, Xinxin Niu:Capacity of blockchain based Internet-of-Things: Testbed and analysis. Internet Things 8 (2019).

[87] Miles Carlsten: The impact of transaction fees on bitcoin mining strategies. Ph.D. thesis, Princeton University, 2016.

[88] Han-Min Kim, Gee-Woo Bock, Gunwoong Lee: Predicting Ethereum Prices using Machine Learning and Block Chain Information. AMCIS 2019 [J]. 2019.

[89] Sean McNally, Jason Roche, Simon Caton: Predicting the Price of Bitcoin Using Machine Learning. PDP 2018: 339-343.

[90] Fang Chen, Hong Wan, Hua Cai, Guang Cheng: Machine Learning in/for Blockchain: Future and Challenges. CoRR abs/1909.06189 (2019).

[91] Shaofei Xue, Hui Jiang, Li-Rong Dai, Qingfeng Liu: Speaker Adaptation of Hybrid NN/HMM Model for Speech Recognition Based on Singular Value Decomposition. Signal Processing Systems 82(2): 175-185 (2016).

[92] Dima Suleiman, Arafat Awajan, Wael Al Etaiwi: The Use of Hidden Markov Model in Natural ARABIC Language Processing: a survey. EUSPN/ICTH 2017: 240-247.

[93] Haitao Zhou, Jin Chen, Guangming Dong, Ran Wang: Detection and diagnosis of bearing faults using shift-invariant dictionary learning and hidden Markov model[J]. Mechanical systems and signal processing, 2016, 72: 65-79.

[94] Eun-chong Kim, Han-wook Jeong, Nak-young Lee: Global Asset Allocation Strategy Using a Hidden Markov Model[J]. Journal of Risk and Financial Management, 2019, 12(4): 168.

[95] Iman Abu Hashish, Fabio Forni, Gianluca Andreotti, Tullio Facchinetti, Shiva Darjani: A Hybrid Model for Bitcoin Prices Prediction using Hidden Markov Models and Optimized LSTM Networks. ETFA 2019: 721-728.

[96] Constandina Koki, Stefanos Leonardos, Georgios Piliouras: Do Cryptocurrency Prices Camouflage Latent Economic Effects? A Bayesian Hidden Markov Approach[C]//Multidisciplinary Digital Publishing Institute Proceedings. 2019, 28(1): 5.

[97] Constandina Koki, Loukia Meligkotsidou, Ioannis Vrontos: Forecasting under model uncertainty:Non-homogeneous hidden Markov models with Polya-Gamma data augmentation. arXiv 2019, arXiv:1802.02825.

[98] Ross C. Phillips, Denise Gorse: Predicting cryptocurrency price bubbles using social media data and epidemic modelling. SSCI 2017: 1-7.

[99] David García, Claudio Juan Tessone, Pavlin Mavrodiev, Nicolas Perony:The digital traces of bubbles: feedback cycles between socio-economic signals in the Bitcoin economy. CoRR abs/1408.1494 (2014).

[100] Kwansoo Kim, Sang-Yong Tom Lee, Said ASSAR: Coin market behavior using social sentiment Markov chains[C]. 2019.

[101] Daniel Chappell: Regime heteroskedasticity in Bitcoin: A comparison of Markov switching models[J]. Available at SSRN 3290603, 2018.

[102] Franc Klaassen: Improving GARCH volatility forecasts with regime-switching GARCH[J]. Empirical Economics, 2002, 27(2):363-394.

[103] Paulo Vitor Jordão da Gama Silva, Marcelo Klotzle, Antônio Carlos Figueiredo Pinto, Leonardo Lima Gomes: Volatility estimation for cryptocurrencies using Markov-switching GARCH models[J]. International Journal of Financial Markets and Derivatives, 2019, 7(1): 1-14.

[104] Miriam Sosa, Edgar Ortiz, Alejandra Cabello: Bitcoin Conditional Volatility: GARCH Extensions and Markov Switching Approach[J]. Disruptive Innovation in Business and Finance in the Digital World (International Finance Review, Vol. 20), Emerald Publishing Limited, 2019: 201-219.